\documentstyle[12pt,aaspp4]{article}
\begin{document}

\title{GAMMA-RAY BURSTS AS HYPERNOVAE}

\author{Bohdan Paczy\'nski}
\affil{Princeton University Observatory, Princeton, NJ 08544--1001, USA}
\affil{e-mail: bp@astro.princeton.edu}

\begin{abstract}

The energetics of optical and radio afterglows following BeppoSAX and BATSE
gamma-ray bursts (GRBs) suggests that gamma-ray emission is not 
narrowly collimated, but a moderate beaming is possible, so the
total energy of a GRB may be in the range $ \sim 10^{50} - 10^{51} $ erg.
All attempts to generate a fireball powered by neutrino-antineutrino
annihilation have failed so far, and a rapid rotation combined
with a magnetic field of $ \sim 10^{15} $ gauss gains popularity.

In this paper a hypernova scenario is described: a
collapse of a massive member in a close binary system, similar to the 
`failed' Type Ib supernova model proposed by Woosley (1993).
The collapse may lead to explosion, with energy
transmitted from the rapidly spinning hot neutron core or black hole
to the envelope by a strong magnetic field, as in a supernova model 
proposed by Ostriker and Gunn (1971).  However, because of a large
mass and rapid rotation of the core the explosion of
a hypernova may release up to $ \sim 10^{54} $ erg
of kinetic energy, creating a `dirty' fireball.

In this scenario a moderately non-spherical explosion may accelerate a very
small fraction of matter to a very large Lorentz factor, and this may
give rise to a gamma-ray burst and its afterglow, just like in a
conventional fireball model.  However, the highest velocity ejecta from
a hypernova are followed with matter which expands less rapidly
but carries the bulk of kinetic energy, providing a long term power source
for the afterglow.  If the afterglows remain luminous for a very long time
then the proposed hypernova scenario may provide an explanation.
Pre-hypernovae, being massive stars, are likely to be located in the
young, star forming regions.

\end{abstract}


%\section 1
\section{Introduction}

The recent detection of gamma ray bursts by BeppoSAX: 970228 (Costa et al.
1997a), 970402 (Feroci et al. 1997, Heise et al. 1997), and 970508
(Costa et al. 1997c), and BATSE (970616: Connaughton et al. 1997),
and the corresponding afterglows in X-rays
(970228: Costa et al. 1997b,d; 970402: Piro et al. 1997a; 970508: Piro et al.
1997b; 970616: Marshall et al. 1997), 
in optical (970228: Groot et al. 1997, van Paradijs et al. 1997,
Sahu et al. 1997, Galama et al. 1997; 970508: Bond 1997), and
in radio domains (970508: Frail et al. 1997), has lead to the first
direct estimate of the distance to one burst (970508), with
its redshift in the range $ 0.8 < z < 2.3 $ (Metzger et al. 1997a),
and its total energy at $ \sim 10^{52} $ erg (assuming spherical
emission, Waxman 1997b).
The afterglows are thought to be due to collision between the GRB ejecta and
the surrounding medium, be it circumstellar, interstellar, or intergalactic.

In the following section the essentials of the currently popular GRB
and afterglow
models are outlined, with the emphasis on the underlying assumptions and
the major unresolved issues.  

In the third section a schematic hypernova scenario is presented.
It is based on many unrelated papers written by many authors.
Colgate (1968), five years prior to the discovery of gamma-ray bursts
(Klebesadel et al. 1973), proposed that supernovae may give rise
to gamma-ray bursts when the shock breaks out through the stellar
surface.  Two related concepts are important.  First,
a strong, adiabatic shock can accelerate small amount of matter
to relativistic velocities if the density of ambient medium falls
rapidly enough with distance.  Second, the highest velocity ejecta
are followed by the bulk of mass expanding less rapidly but carrying
most of kinetic energy.  Ostriker and Gunn (1971) proposed
that pulsars, rapidly rotating magnetized neutron stars, may power
supernova explosions.  The important concept is the magnetic energy
transfer from the collapsed, rapidly rotating  stellar core to the envelope.
Woosley (1993) proposed a `failed' Type Ib supernova model to explain
gamma-ray bursts at cosmological distances.  The important concept is
that a massive, collapsed, rapidly rotating stellar core may form a
black hole of $ \sim 5 - 10 ~ {\rm M}_{\odot} $ with a massive high
density disk spinning around it.  Iben \& Tutukov (1996) pointed out
that rapidly rotating magnetized neutron stars, i.e. radio pulsars,
are likely to form in close binaries only.  The important concept
is the need of a close binary companion to keep a pre-supernova star
rapidly rotating.  Our hypernova scenario combines all five
concepts listed above.

Finally, some observational tests of the hypernova scenario, and
some specific problems to be modeled theoretically
are discussed in the fourth section.

%\section 2
\section{A Fireball Model}

In a standard afterglow model energy of the order $ 10^{51} $ erg
is deposited in a small volume and mixed with a small amount of
matter, $ \sim 10^{-6} ~ {\rm M}_{\odot} $, generating an ultra-relativistic
outflow with a bulk Lorentz factor $ \sim 300 $.  A collision of
the ejecta with ambient medium, be it circumstellar, interstellar,
or intergalactic gives rise to a gamma-ray burst (Rees \& M\'esz\'aros
1992), and to a subsequent afterglow (Paczy\'nski \& Rhoads 1993,
Wijers et al. 1997, Waxman 1997a,b, Sari 1997, Vietri 1997, and references 
therein).

In the simplest model GRB ejecta expand as a thin spherical shell.
The shocks formed by the shell interaction with ambient
medium slow down the expansion, with the bulk Lorentz factor dropping
down to $ \sim 3 $ by the time a significant radio power is generated.
The overall energetics is in a rough agreement with the observations,
with synchrotron radiation dominating the X-ray, optical and radio
emission of the afterglows.  An extreme beaming of the ejecta and
the gamma-ray emission seems to be ruled out by the data
(Rhoads 1997), though a moderate beaming is certainly possible.

The most fundamental problem, the ultimate energy source and the
physical process leading to fireball formation, has not been
solved yet.  In particular,
once popular models based on neutrino-antineutrino annihilation
fail by several orders of magnitude (Jaroszy\'nski 1996, 
Janka \& Ruffert 1996, Vietri 1996, Ruffert et al. 1997,
M\'esz\'aros \& Rees 1997, and references therein).
Therefore, very strong magneting fields ($ B \sim 10^{15} $ gauss)
combined with rotation are now suggested as a new way to create and
power a fireball (Vietri 1996, M\'esz\'aros \& Rees 1997).

The next fundamental issue
is the lifetime of the ultimate energy source: is it impulsive (Rees
\& M\'esz\'aros 1992, Wijers, Rees \& M\'esz\'aros 1997), or is the
source active for the duration of the GRB, i.e. up to a several minutes,
or even longer
(Paczy\'nski 1991, 1993, Waxman 1997a,b, and references therein).  
In the former case the GRB's rapid variability must
be due to a complicated structure of the so called `external shock'
formed by the ejecta colliding with ambient medium, while in the
latter case variability is caused by `internal shocks' within the
ejecta (Paczy\'nski 1991, Paczy\'nski \& Xu 1994, Rees \& M\'esz\'aros 1994,
Waxman 1997a,b, Sari 1997, Vietri 1997, and references therein).
Recently Sari et al. (1996)
and Sari \& Piran (1997) presented strong arguments in favor of the
internal shock scenario, by demonstrating that under reasonable
assumptions an external shock cannot provide rapid and large amplitude
variability commonly observed in GRBs.  The afterglow
is believed to be caused by the external shock.  Therefore, the issue
of internal vs external shocks for the GRB emission is closely related
to the question: what is the distinction, if any, between the burst itself 
and the aftergflow?

A standard model of the afterglow
assumes that the magnetic field, the electrons and the protons have
energy densities close to equipartition, as this is required for
the efficient synchrotron emission.  Yet, a simple shock compression
is not sufficient to establish anything even close to
equipartition, and a strong turbulence driven by some instability
is always assumed, though the assumption is usually implicit.
The complexity of this process and the possibility that most GRB
energy is released in concentrated `hot spots', like solar flares,
weakens the Sari \& Piran's (1997) case against the
external shock variability (Goodman 1997).  In my view the issue: is GRB 
phenomenon due to internal or external shocks has not been
resolved yet.  

The degree of GRB beaming is also a critical issue as the required energetics
and the rate of bursts depend on it.  Fortunately, this question will
likely be answered soon with the observations of afterglows related and not 
related to known GRBs (Rhoads 1997).   An extreme beaming is already excluded
by a reasonable agreement between the current spherical models and the
observations of the afterglows.

%\section 3
\section{A Hypernova Scenario for GRBs}

The original work by Colgate (1968) preceded the discovery of gamma-ray
bursts by Klebesadel et al. (1973).  In Colgate's model a spherical
shock developed in a supernova explosion accelerated a small fraction
of outer stellar layers to highly relativistic velocities, and
gave rise to a gamma-ray burst.  Following 
the discovery of actual bursts Colgate (1974) estimated their distance to 
be $ \sim 10 $ Mpc as the detected fluence was $ \sim 10^{-4} ~
\rm erg ~ cm^{-2} $ and his model was supposed to generate $ \sim 10^{48} $
erg in gamma-rays.  Unfortunately,
careful modeling of the shock breakout demonstrated that radiative heat
losses damp out the shock when it comes close to the stellar surface,
and there is no reason to expect that a bulk ultrarelativistic flow
and gamma-ray emission
develop (Ensman \& Burrows 1992, and references therein).

Even though the original Colgate's model does not work, and even in
its most optimistic formulation it was not claimed to provide 
$ \sim 10^{51} $ erg in gamma-rays, it contains a very important
concept: only a very small fraction of explosion ejecta reaches
relativistic velocity.  If this concept is applied to actual GRBs
then the bulk of mass and kinetic energy of explosion may be in
relatively slowly moving ejecta, and the total energy involved may be
a few orders of magnitude larger than that radiated away in
gamma-rays.  This would have a profound effect not on the GRB itself, 
but on its afterglow: when the leading shock slows down by sweeping
up ambient medium, the slower moving matter far behind the shock
can catch up, and provide a powerful and long lasting energy source
for the afterglow.

The total amount of energy required for GRB 970508
is $ \sim 10^{52} $ erg, assuming spherical
symmetry (Waxman 1997b).  A moderate amount of beaming, still compatible
with the current data (Rhoads 1997) may reduce this
demand to $ \sim 10^{50}-10^{51} $ erg.  The total kinetic energy
of either Type I or Type II supernova is $ \sim 10^{51} $ erg, and
our expectation is that only a small fraction of the kinetic energy
of any explosion
may be converted into gamma-rays.  A natural solution of the apparent
energy crisis would be provided by a much more powerful event,
with $ 10^{53} - 10^{54} $ erg of kinetic energy,
which we call a hypernova.  A similar term: `radio hypernova
remnant' was proposed by Wilkinson \& de Bruyn (1990) to describe
objects which were $ \sim 100 $ times brighter than ordinary.

Super-strong $ \sim 10^{15} $ gauss magnetic fields were 
suggested some time ago (Narayan et al. 1992, Paczy\'nski 1993)
as possibly relevant to the generation of GRBs.
Recently super-strong magnetic fields were 
proposed again to mediate efficient energy transfer from 
the orbital energy of a pair of merging neutron star (Vietri 1996), or
the spin energy of a stellar mass black hole (M\'esz\'aros \& Rees 1997),
into a fireball needed to power cosmic gamma-ray bursts.  
Such strong magnetic fields may be created either
by a dynamo (Balbus \& Hawley 1991,
Duncan \& Thompson 1992, and references therein), or they
may be fossils from the time when the star was on the main sequence.
Note that main sequence magnetic Ap stars have fields up to
$ 20,000 $ gauss (Borra et al. 1982, and references therein).
A core contraction of a magnetic Ap star (with flux conservation)
may give rise to a $ 10^8 - 10^9 $ gauss field in a white dwarf, and 
indeed some white dwarfs are known to have such fields
(Chanmugam 1992, and references therein).
The broad lines of O-type stars prevent measurements
of their magnetic fields, but there is no reason for strong fields not to
be present on some O-type stars.  When a core of a massive
star collapses to form a neutron star or a black hole (conserving
magnetic flux) then the field may reach $ \sim 10^{15} $ gauss.
Two independent lines of evidence suggest that a neutron star underlying 
the famous March 5th 1979 burst and its repeaters
has a field of $ \sim 5 \times 10^{14} $ gauss (Duncan \& Thompson 1992,
Paczy\'nski 1992).  Radio pulsars have fields up to $ \sim 2 \times 10^{13} $
gauss (Taylor \& Stinebring 1986, and references therein).

The idea that a strong magnetic field can transmit energy out from a
rapidly spinning neutron core and give rise to a supernova 
was proposed a long time ago by Ostriker \& Gunn (1971).  However,
with the $ \sim 10^{12} $ gauss field envisioned at the time the
energy transfer was not likely to be rapid enough for an explosion.
This is the reason to invoke a field of $ \sim 10^{15} $ gauss.
The maximum rate at which magnetic field can extract energy from
a relativistically spinning object, like a black hole, can be
estimated to be
$$
L_{B,max} \approx B^2 R^2 c \approx
10^{52} ~ {\rm [ erg ~ s^{-1}]} ~ \left( { B \over 10^{15} } \right) ^2
~ \left( { R \over 10^6 ~ cm } \right) ^2 ,      \eqno(1)
$$
following Znajek (1977).  

The known stellar mass black hole candidates have masses 
$ \sim 5 - 10 ~ {\rm M}_{\odot} $ (Cowley 1992), and right after
their formation in a core collapse of a massive star they may have up
to $ 5 \times 10^{54} $ erg stored in their rotation.  If a large
fraction of that
energy is extracted rapidly with a strong enough magnetic field,
and not wasted as neutrino emission, then the kinetic energy of explosion 
may be $ E_{kin} \sim 10^{54} $, perhaps even more.
If the total ejected mass is $ M_{env} \sim 10 ~ {\rm M}_{\odot} $
then all ejecta would be subrelativistic, with the average expansion
velocity 
$$
v_{av} \approx \left( { E_{kin} \over 2 M_{env} } \right) ^{1/2}
\approx 10^{10} ~ {\rm cm ~ s^{-1}} ~
\left( { E_{kin} \over 10^{54} ~ {\rm erg}} ~ 
{ 10 ~ M_{\odot} \over M_{env} } \right) ^{1/2} .
\eqno(2)
$$
A small fraction of the envelope mass my be accelerated to a very large
Lorentz factor by the explosion process within the star, or a strong shock
expanding rapidly into circumstellar envelope with a steep
density gradient.  These ultra-relativistic ejecta provide the conditions
required for a fireball model of gamma-ray bursts.

It seems likely that a combination of a superstrong magnetic field and
rapid rotation may exist in close binaries only, where the tides
keep the stars in co-rotation with the short period orbital motion 
(Iben \& Tutukov 1996).  This 
implies a small orbit and small radii of the two
components, which are likely to be Wolf-Rayet stars, like the
mass losing component of Cyg X-3 (van Kerkwijk et al. 1996).
This has two consequences.  First, the initial explosion may
produce very little optical emission as the exploding star
is so compact.  Second, there is likely to be an extended
circumstellar envelope created with the stellar wind preceding
the explosion.  The propagation of sub-relativistic ejecta through
this envelope may accelerate small fraction of matter to 
a very large Lorentz factor if the density is falling rapidly
enough with radius and the shock becomes collisionless and
more or less adiabatic.

An explosion of a rapidly spinning star is likely to produce some
departure from the spherical symmetry, with jet-like structure forming
along the rotation axis.  Somewhat similar suggestion to the
one presented here were made in the past (e.g. Paczy\'nski 1991,
Narayan et al. 1992, Paczy\'nski 1993, Woosley 1993, M\'esz\'aeos
\& Rees 1997).  A less explosive and longer lasting event was referred 
to as a microquasar, by analogy with the popular quasar models.

The following is a schematic outline of a hypernova event.

We begin with a pair of massive stars in a close binary orbit, which
is a standard scenario for the subsequent formation of massive X-ray
binaries (Canal et al. 1990, Iben \& Tutukov 1996, and references therein).
The presence of the companion star is needed just to keep the massive star
rapidly rotating, which is done best by coupling it tidally to the binary
orbit.  In order to maintain rapid rotation at the
end of nuclear evolution the binary period has to be short, i.e. the
orbit has to be compact, and therefore hydrogen envelope had to be lost.
A helium star
of $ \sim 20 ~ {\rm M}_{\odot} $ has a radius of $ \sim 1.3 ~ R_{\odot} $.
Two such stars can comfortably fit in a binary with a separation
$ {\rm A \sim 5 ~ R_{\odot}} $, i.e. the binary period of 5 hours.
Of course, the binary had to be more massive and the orbit much larger
at the time when the two stars were on the main sequence.  However,
it is common for binaries to loose mass and angular momentum during
the evolution.  A helium star of $ \sim 20 ~ {\rm M}_{\odot} $ develops
an iron core of $ \sim 2.1 ~ {\rm M}_{\odot} $ (Woosley \& Weaver 1986,
and references therein).

At this stage the evolution follows the scenario proposed by Woosley (1993).
The hot and rapidly spinning neutron core formed in a gravitational collapse
cools off by neutrino emission and becomes a stellar mass black hole.  The
stellar mantle collapses into a hot neutron torus spinning around the
black hole, and gradually accreting into it.  Here the `failed' Type Ib
supernova scenario of Woosley (1993) and the hypernova scenario diverge.
Model calculations (Jaroszy\'nski 1996, Janka \& Ruffert 1996, Ruffert et al.
1997) demonstrated that powerful streams of neutrinos and antineutrinos 
fail to generate a fireball.  Therefore I propose that a magnetic field
of $ \sim 10^{15} $ gauss, like that suggested by Vietri (1996) and
M\'esz\'aros \& Rees (1997), provides efficient coupling between the
spinning black hole and torus and the stellar envelope.  This is similar to
the pulsar driven supernova model proposed by Ostriker \& Gunn (1971).
In our scenario the magnetic field is stronger than a typical pulsar's field 
by a factor $ \sim 10^3 $, the efficiency of magnetic energy transport is
higher by a factor $ \sim 10^6 $, and it may be sufficient to explode the 
stellar envelope.  With the collapsed core a few times more massive than a 
neutron star, and rotation rate close to critical, the available energy 
is much higher too, equivalent to a fair fraction of the core rest mass,
i.e. up to $ 10^{54} $ erg, or even more.  As the explosion
is expected to be much more energetic than a supernova, a name 
hypernova is proposed.

There are several differences between the `failed' Type Ia supernova
and the hypernova scenarios.  Not only neutrino energy transport is replaced
with magnetic transport.  The rapid rotation, which was somewhat
ad hoc in Woosley's model is a direct consequence of the close binary
nature of the massive star.  Also, a hypernova does not have to create
a fireball, a strong explosion may be adequate.  
With $ \sim 10 ~ {\rm M}_{\odot} $
of the envelope mass ejected with the energy of $ \sim 10^{54} $ erg,
the average velocity of ejecta is $ \sim 10^{10} ~ {\rm cm ~ s^{-1}} $.
The rapid rotation is likely to make the expansion easier along the
rotation axis.  Therefore, moderately jet-like ejection of a fraction of the
envelope mass with moderately relativistic velocity is likely.  Any massive
star has a strong wind throughout its evolution, and the hypernova ejecta
will produce a relativistic shock in the circumstellar medium.  At low
enough density the shock will be collisionless and adiabatic.  At steep
enough density gradient the shock may accelerate a small amount of
mass to a very large Lorentz factor.  Therefore, a hypernova may succeed
where Colgate's supernova failed, and with large enough Lorentz factor
of the shock it may generate a gamma-ray bursts in a way similar to
the popular fireball model.

%\section 4
\section{Discussion}

The most dramatic difference between a hypernova and a fireball
scenarios is in the long term behavior of the afterglows.

In the currently popular fireball model 
$ \sim 10^{51} $ erg of energy is released in a small volume, with only
a very small amount of matter present there, $ \sim 10^{-6} ~ {\rm M}_{\odot}
\approx 3 \times 10^{27} ~ {\rm g} $.  The expansion with a very large
bulk Lorentz factor sets in immediately, and the ejecta form a very narrow 
shell (Rees \& M\'esz\'aros 1992), 
with the interior of the rapidly growing sphere being almost
empty.  This has a profound impact on the afterglow's energetics,
which is powered by the thin shell only.  In the hypernova scenario
proposed in this paper the explosion establishes a massive
outflow with only a very small fraction of the total mass accelerated
to the maximum Lorentz factor.  The interior of the fastest shell is
full of matter expanding with a lower velocity.
It follows that when the shock is decelerated by the
ambient medium of more or less constant density,
the matter expanding behind it will catch up, providing a fresh
supply of kinetic energy to power the afterglow emission.
This is the single most important difference between the hypernova
and the fireball: in a hypernova model the afterglow should last for
a much longer time.

Shock acceleration of some ejecta to a very large Lorentz factor
is a possible mechanism.  It is also possible that the explosion
of the rapidly rotating envelope becomes bipolar, and forms
a highly collimated and ultra-relativistic jet along the spin
axis.  In this case the afterglow would be powered not only by
the GRB jet, but also by the less rapidly expanding matter, located
farther from the axis.  The effect would be to provide extra
energy for the afterglow, and therefore to extend its lifetime.

I expect that future observations of the aftergflows
will either support or eliminate the hypernova model.  If afterglows
remain luminous far too long to be explained with a
fireball model than their long term behavior will provide information
about the properties of hypernovae.  Of course, the specific mechanism
for the explosion proposed in this paper, i.e. efficient magnetic transfer
of energy from the rapidly rotating, collapsed massive core may turn
out to be wrong, but some other means to power hypernovae may be found.
As the pre-hypernova is required to be a massive star, it is likely
to be located in a region of vigorous star formation.  Therefore, it is 
interesting that the afterglow of GRB 970508 has emission lines in its 
optical spectrum corresponding to the redshift z = 0.835 (Metzger 1997b).

Note, that while a hypernova scenario places GRBs in star forming regions,
any merging neutron star scenario places them far away from such regions.
By the time the binary merges it has move a long way from
its place of origin, because il must acquire a large velocity
as a consequence of two consecutive supernovae explosions, even if those
explosions were spherically symmetric.  Therefore, the location of
gamma-ray bursts will reveal which scenario is correct.

Future observations of the afterglows related to gamma-ray bursts will
provide information about the energetics.  Here the distinction between
a hypernova and a fireball scenarios may not be as clean as one might
wish.  The simplest clean fireball model puts all ejecta in a thin shell,
but it is done so by design: the initial energy per baryon is assumed
to be constant.  If the initial fireball is more diverse, with a broad
range of energies per baryon then there may be a continuum of possible
scenarios filling the full range of possibilities, between a clean simple
fireball and a hypernova.  

A detection of afterglows
not related to GRBs will provide information about the beaming of gamma-ray
emission (Rhoads 1997). 

The proposed scenario is made of many components, some robust, and
some speculative.  There is nothing speculative about the existence of
short period massive binaries (Canal et al. 1990, Iben \& Tutukov 1996,
and references therein).  The collapse of a rapidly rotating massive core
to form a stellar mass black hole is a logical step in the evolution
(Woosley 1993, and references therein).  The efficient magnetic energy
transport from the core to the envelope, and an explosion with 
$ \sim 10^{54} $ erg of kinetic energy is a speculation.  Of course,
similar process has been proposed many times by many authors in various
contexts (Ostriker \& Gunn 1971, Blandford \& Znajek 1977, M\'esz\'aros
\& Rees 1997).  However, practical feasibility has neither been proven
nor disproven.  The next step, the acceleration of a small fraction
of ejecta to ultra-relativistic velocity is somewhat speculative,
though shock acceleration and/or the formation of a narrow jet seem to
be reasonable.  The final step, the gamma-ray emission and the afterglow
have physics almost identical with that of the popular fireball models.

A hypernova or a micro-quasar scenario is very schematic, and the diversity
of possible developments is likely to be very rich.  No firm conclusions
are possible without detailed and complicated numerical models.
If the history of Type II supernovae and the
success (or lack of success) of models to predict the
effectiveness of the `bounce' are to be a guide then we may not expect
a clear and robust answer to most hypernova related
questions for a few more decades.
However, a semi-empirical approach may be possible with the future
studies of GRB afterglows.

\acknowledgments{It is a great pleasure to
acknowledge many stimulating discussions and useful comments by
J. Goodman, R. Chevalier, M. Rupen, S. van den Bergh, M. Vietri, E. Waxman,
S. E. Woosley, and many participants of the morning coffees at Peyton Hall.
This work was supported with the NSF grants AST--9313620 and AST--9530478.}  

%\newpage

%REFERENCES

\end{document}